\documentclass[aps,prb,twocolumn,superscriptaddress,showpacs]{revtex4}
\usepackage{graphicx}
\usepackage{dcolumn}
\usepackage{hyperref}

\begin{document}

\title{Acoustic vibrations of anisotropic nanoparticles}

\author{Lucien Saviot}
\affiliation{Institut Carnot de Bourgogne,
UMR 5209 CNRS - Universit\'e de Bourgogne,
9 avenue A. Savary, BP 47870, F-21078 Dijon Cedex, France}
\email{lucien.saviot@u-bourgogne.fr}

\author{Daniel B. Murray}
\affiliation{Mathematics, Statistics and Physics Unit,
The University of British Columbia Okanagan, 3333 University Way,
Kelowna, British Columbia, Canada V1V 1V7}
\email{daniel.murray@ubc.ca}

\date{\today}

\begin{abstract}

Acoustic vibrations of nanoparticles made of materials with
anisotropic elasticity and nanoparticles with non-spherical
shapes are theoretically investigated using a homogeneous
continuum model. Cubic, hexagonal and tetragonal symmetries of
the elasticity are discussed, as are spheroidal, cuboctahedral
and truncated cuboctahedral shapes.  Tools are described to
classify the different vibrations and for example help identify
the modes having a significant low-frequency Raman scattering
cross-section.  Continuous evolutions of the modes starting from
those of an isotropic sphere coupled with the determination of
the irreducible representation of the branches permit some
qualitative statements to be made about the nature of various
modes.  For spherical nanoparticles, a more accurate picture is
obtained through projections onto the vibrations of an isotropic
sphere.

\end{abstract}

\maketitle

\section{Introduction}

The lowest frequency vibrations of isolated nanoparticles
are in the THz range, on the order of the speed of sound divided by the
dimension. These are commonly refered to as confined acoustic phonons
and are unrelated to optical phonons.
There have been numerous experimental and theoretical studies on the
acoustic vibrations of nanoparticles in the last few decades. These
vibrations have been observed by
a variety of experimental techniques including
low frequency Raman scattering,\cite{Duval86} 
time resolved femtosecond pump-probe
experiments,\cite{DelFattiJCP1999,IkezawaJPSJ05,BurginNL08}
infrared absorption,\cite{MurrayJNO06,LiuAPL08}
inelastic neutron scattering\cite{SaviotPRB08} and
persistent spectral hole burning.\cite{IkezawaJPSJ05}

Reasonable estimates for the mode frequencies are obtained using
the 1882 Lamb solution of the continuum elastic problem for an
elastically isotropic, homogeneous, free sphere.\cite{lamb1882}
This provided sufficiently good agreement to confirm that confined
acoustic phonon modes were really being observed.
However, this model was unable to deal with the anisotropy of
actual samples.

Recent advances have permitted the creation of high quality elastically
anisotropic samples.\cite{PortalesPNAS08} The essential features are (1)
a narrow size distribution (2) good crystallinity so that a significant
amount of nanoparticles in the sample are
mono-domain (3) controlled
shape of nanoparticles and (4) separation of nanoparticles so that
they vibrate as independent units. As a result, the vibrational modes
of elastically anisotropic nanoparticles have been observed. This has
created the need for an alternative to the Lamb model capable
of dealing with nanoparticles with lower symmetry.

In this work, we use the method of Visscher
\textit{et al.}\cite{visscher} which is a standard numerical
approach suitable for the calculation of the frequencies and the
wavefunctions of the vibrations of such nanoparticles. The
symmetry of these modes,
their volume variation and their Lamb mode parentage
are determined and applied to the prediction of their observation by
different experimental techniques such as inelastic light scattering.

\section{Methods}

The situation for an isotropic nanoparticle will now be summarized.  In
this case, the system is spherically symmetric. Thus, vibrational modes
can be classified by their angular momentum number $\ell \ge 0$ and its
$z$-component $m$. Modes can also be classified either as torsional (T)
or spheroidal (S). Finally, modes are also indexed in order of frequency
by $n \ge 1$ ($n=1$ corresponds to the first harmonic (fundamental mode),
$n=2$ to the second harmonic and so on). In the following, we will
indicate Lamb modes using the compact notation X$_{\ell m}^n$ where X=S
or X=T.

All modes can be observed by inelastic neutron scattering in the
typical situation where the wavelength of the neutrons is much
smaller than the nanoparticle size. For nanoparticles whose
dimension is small compared to the wavelength of light (dipolar
approximation), Raman only detects S$_{0}$ and S$_2$, infrared
absorption only
detects S$_1$\cite{duval92} and time-resolved femtosecond
pump-probe experiments typically only detect S$_0$. In the
following, we will assume that the nanoparticles are small
enough so that the dipolar approximation holds.

Nanoparticles with either isotropic or anisotropic elasticity
will be considered in the following. Isotropic elasticity is
considered mainly for comparison with previous studies. 
Anisotropic elasticity is
used for perfect nanocrystals consisting of a single domain.
We will refer to such nanoparticles as being ``mono-domain'' in
the following. In a small nanoparticle, a mono-domain structure
is not necessarily energetically favorable. For example, it is
well known that multiply-twinned silver nanoparticles are much
more stable for certain ranges of size.\cite{InoJPSJ69}

\subsection{Calculation of frequencies and associated displacements}

The frequencies and their associated displacements for an
anisotropic nanoparticle have been calculated using the approach
introduced by Visscher \textit{et al.}\cite{visscher} which
also assumes continuum elasticity. Other authors have already
confirmed that the convergence of this method is faster than the
convergence of finite element methods at least in some
cases.\cite{HeyligerJSV08}
The relevance of continuum
elasticity for nanoparticles has been confirmed using atomistic
calculations\cite{ChengPRB05,ErratumChengPRB05,CombePRB07,RamirezJASA08}
for nanoparticles larger than 2-3~nm and even for ZnO nanoparticles for
which surface relaxation and stress are significant.\cite{CombePRB09}
Most of the results presented in this paper have been obtained for
nanospheres whose diameter is 10~nm which is well above these limits.
It is possible to extrapolate them to different sizes since the
frequencies vary as the inverse diameter. However care should be taken
not to consider very small nanoparticles for which surface effects could
significantly alter the validity of the continuum approximation.
The calculational method for the modes gives each mode in terms of
power series coefficients $a_{ijpqr}$ so that:

\begin{equation}
\vec u_i(x,y,z) = \sum_{pqr} \sum_{j=x,y,z}
  a_{ijpqr}
  x^p y^q z^r \vec j
\end{equation}

The power expansion covered $0 \le p+q+r \le 20$ for
good convergence for all the modes we are interested in. The
frequencies for the isotropic spherical case were reproduced with
very good accuracy. The convergence for strongly anisotropic
systems is harder to check. Despite checking that the
frequencies do not significantly change when adding more terms to
the power expansion, we also compared the calculated frequencies
with the Finite Element Mesh Sequence method introduced in a
previous work.\cite{SaviotPRB04}

In this work, all the displacements have been normalized
according to equation~\ref{norm} where $V$ is the volume of the
nanoparticle.\cite{MurrayPRB04}

\begin{equation}
  \int\hspace{-.8em}\int\hspace{-.8em}\int_V \vec u_i \cdot \vec u_j d^3\vec R
  = V \delta_{i j}
\label{norm}
\end{equation}

\subsection{Group theory}

\subsubsection{Degeneracy lifting}

In the absence of spherical symmetry, modes are classified according
to their remaining symmetry. For example, for a spherical
nanoparticle with cubic elasticity, such as Ag, Au and Si, the system
is symmetric under the symmetry operations of a cube, which is the
48-element group $O_h$.

Our calculations return a large number of modes
which have to be considered to interpret inelastic light
scattering spectra or other experimental results. It is
important to use not only the frequencies but also the
wavefunctions in order to do that. Group theory is a very
valuable tool in this context as it allows for example to
identify the Raman active modes and therefore simplify the
assignment process. In this work, we will only consider
nanoparticles whose dimensions are small compared to the
wavelength of light (dipolar approximation) in order to discuss
the selection rules of Raman scattering. Table~\ref{degeneracy}
shows how the degeneracy of the Raman and infrared active modes
of an isotropic sphere\cite{duval92} is lifted or not when
lowering the symmetry. Only point groups relevant for the rest
of this paper are considered.

\begin{table*}
\begin{tabular}{c|c|c|c|c|c}
point group
 & S$_0$
 & S$_1$
 & S$_2$
 & S$_3$
 & S$_4$\\
\hline
O$_{\text{h}}$
 & A$_{\text{1g}}$
 & T$_{\text{1u}}$
 & E$_{\text{g}}$ + T$_{\text{2g}}$
 & A$_{\text{2u}}$ + T$_{\text{1u}}$ + T$_{\text{2u}}$
 & A$_{\text{1g}}$ + E$_{\text{g}}$ + T$_{\text{1g}}$ + T$_{\text{2g}}$\\
D$_{\text{4h}}$
 & A$_{\text{1g}}$
 & A$_{\text{2u}}$ + E$_{\text{u}}$
 & A$_{\text{1g}}$ + B$_{\text{1g}}$ + B$_{\text{2g}}$ + E$_{\text{g}}$
 & A$_{\text{2u}}$ + B$_{\text{1u}}$ + B$_{\text{2u}}$ + 2 E$_{\text{u}}$
 & 2 A$_{\text{1g}}$ + A$_{\text{2g}}$ + B$_{\text{1g}}$ + B$_{\text{2g}}$ + 2 E$_{\text{g}}$\\
D$_{\text{6h}}$
 & A$_{\text{1g}}$
 & A$_{\text{2u}}$ + E$_{\text{1u}}$
 & A$_{\text{1g}}$ + E$_{\text{1g}}$ + E$_{\text{2g}}$
 & A$_{\text{2u}}$ + B$_{\text{1u}}$ + B$_{\text{2u}}$ + E$_{\text{1u}}$ + E$_{\text{2u}}$
 & A$_{\text{1g}}$ + B$_{\text{1g}}$ + B$_{\text{2g}}$ + E$_{\text{1g}}$ + 2 E$_{\text{2g}}$\\
D$_{\infty\text{h}}$
 & A$_{\text{1g}}$
 & A$_{\text{1u}}$ + E$_{\text{1u}}$
 & A$_{\text{1g}}$ + E$_{\text{1g}}$ + E$_{\text{2g}}$
 & A$_{\text{1u}}$ + E$_{\text{1u}}$ + E$_{\text{2u}}$ + E$_{\text{3u}}$
 & A$_{\text{1g}}$ + E$_{\text{1g}}$ + E$_{\text{2g}}$ + E$_{\text{3g}}$ + E$_{\text{4g}}$\\
\end{tabular}
\caption{\label{degeneracy}Degeneracy lifting of spheroidal Lamb modes
with $\ell \le 4$ for various symmetries. The degeneracy lifting for
torsional modes (with $\ell > 0$) is obtained by changing the parities
($u \leftrightarrow g$)}
\end{table*}

\subsubsection{Numerical determination of the irreducible representations}

In order to take full advantage of group theory, it is important
to label the different modes with the corresponding irreducible
representation. Sophisticated and specific approaches could be
considered to restrict the calculations to modes having a
well-defined symmetry.  However, we preferred to keep the
numerical approach detailed previously because it is more
general.  We added a few calculation steps to determine the
irreducible representation from the wavefunctions. It turns out
this can be achieved very reliably and without much additional
calculation time.

$\mathcal{S}$ being a symmetry operation of the point group of concern,
and $u_i$ with $i=1, \ldots n$ a full set of
eigenmodes having the same frequency, the character of $\mathcal{S}$ for
this irreducible group of vibrations is:

\begin{equation}
  \xi\left(\mathcal{S}\right) =
   \sum_{i=1}^n
   \int\hspace{-.8em}\int\hspace{-.8em}\int_V
    \mathcal{S}\left(\vec u_i\left(\vec R\right)\right) \cdot
    \vec u_i\left(\mathcal{S}\left(\vec R\right)\right)
    d^3\vec R
\end{equation}

Such integrals can be calculated accurately and quickly when each
Cartesian component of the displacement field is a sum of terms
of the form $x^p y^q z^r$.  By calculating a few well-chosen
characters, it is then straightforward to distinguish all the
different irreducible representations using the character table
of the point group.  The only restriction to apply this procedure
is that the degeneracy must be known
and therefore the convergence must be good. Of course, a more general
approach is require to handle accidental degeneracies.

In the following we detail some additional ways to improve our knowledge
of these vibrations. These are needed since regarding Raman scattering
many modes are labeled as Raman active due to the irreducible
representation they belong to. However there is not necessarily an
efficient coupling mechanism enabling a significant Raman intensity. The
following tools are designed to somewhat address this problem.

\subsection{Volume variation}

The volume of the nanoparticle does not change for every possible
vibration. For isotropic spherical nanoparticles, the volume changes
only for the spheroidal $\ell=0$ vibration. This volume change is also
involved in the time-resolved femtosecond pump-probe
measurements.\cite{DelFattiJCP1999} As a result, it is
interesting to calculate it for all the vibrations.

The volume variation corresponds to the flux of the displacement
through the surface of the particle.  The expression for a
dimensionless volume variation $\delta V$ is given in
equation~\ref{vol}.  Using the divergence theorem, this 2D
integral can be turned into a volume integral involving
derivatives of $x^p y^q z^r$ functions which are already
calculated in the frame of Visscher's method.\cite{visscher}
Therefore this quantity can also be accurately and
efficiently calculated. It can be shown that the volume variation is
different from zero for fully symmetric vibrations only, \textit{i.e.}
for A$_{\text{1g}}$ vibrations for the symmetries considered here.

\begin{equation}
  \delta V_i \cdot V^{2/3} = \left|
  \int\hspace{-.8em}\int_S \vec u_i\left(\vec R\right) \cdot \vec {dS}  \right|
  = \left| \int\hspace{-.8em}\int\hspace{-.8em}\int_V div\left(\vec
u_i\left(\vec R\right)\right) d^3\vec R \right|
\label{vol}
\end{equation}

Of course, the value of $\delta V$ depends on the normalization
(see equation~\ref{norm}).

\subsection{Smooth variation of parameters}
\label{branches}

In order to follow how the vibrations evolve when lowering the
symmetry, it is possible to follow the frequencies of the
different modes while slowly varying the parameters (for example
the shape or the elastic constants of the material the
nanoparticle is made of).  The curve representing the variation
of one frequency is called a branch in the following.  Group
theory can help plotting such branches more reliably because all
the different points on a given branch share the same irreducible
representation. Moreover branches having different irreducible
representations can cross but not branches having the same
irreducible representation. This originates from the coupling
between the different branches being due to the anisotropy itself
whose irreducible representation is A$_{\text{1g}}$ for the point
groups we focus on. As a result, all $n^{\text{th}}$ modes having a given
irreducible representation belong to the same branch and can be safely
connected.

In this work, branches due to varying elastic anisotropy are
calculated by changing the stiffness tensor using
$C(x) = (1-x) C^{\text{iso}} + x C^{\text{ani}}$
with $0 \le x \le 1$, $C^{\text{iso}}$ being the tensor of the
isotropic material (which can be obtained by averaging the sound
velocities\cite{SaviotPRB04} or other
methods\cite{NorrisJMMS06} or using measured longitudinal and transverse
sound velocities) and
$C^{\text{ani}}$ the anisotropic tensor.
The parameters used in this work are given in Ref.~\onlinecite{parameters}.
It should be noted that such branches are made of fictive materials
except for $x=1$ where the real elastic parameters of the bulk material
are used.

\subsection{Projections}

When the lowering of the symmetry is not due to a change in the
shape of the nanoparticle, it is possible to compare the
wavefunctions (omitting their time dependences) of the two
systems.  As commonly done for atomistic calculations\cite{ChengPRB05,ErratumChengPRB05,CombePRB07}, we
calculated the projection of the displacements
of spherical nanoparticles onto Lamb modes. For a
given mode whose displacement is $\vec u_i\left(\vec R\right)$ the
projection onto the Lamb mode $\vec{\text{X}}_{\ell m}^n$ is defined as:

\begin{equation}
P(u_i,\text{X}_{\ell m}^n) = \frac{1}{V}
   \int\hspace{-.8em}\int\hspace{-.8em}\int_V
    \vec u_i\left(\vec R\right) \cdot
    \vec{\text{X}}_{\ell m}^n\left(\vec R\right)
    d^3\vec R
\end{equation}

The orthonormality and completeness of the Lamb modes implies
that, for any $i$,

\begin{equation}
\sum_\ell \sum_m \sum_n P(u_i,\text{X}_{\ell m}^n)^2 = 1
\end{equation}

Likewise, the orthonormality and completeness of the modes of
the anisotropic nanoparticle implies that, for any $\ell$, $m$
and $n$,

\begin{equation}
\sum_i P(u_i,\text{X}_{\ell m}^n)^2 = 1
\end{equation}

However, a more relevant quantity is obtained by summing the
squared projections over all the degenerate Lamb modes
\textit{i.e.} over $m$.  As a result, the total squared
projection of $u_i$ onto the subspace spanned by the
$2 \ell + 1$ modes X$_\ell^n$ is:

\begin{equation}
  \sum_{m=-\ell}^\ell P(u_i,\text{X}_{\ell m}^n)^2
\end{equation}

The Lamb projections given in the fourth columns of tables~\ref{TabAg}
through \ref{TabCo} correspond to this last quantity.
It represents the ratio of the energy contained in the projection
onto Lamb mode X$_{\ell}^n$ to that of mode $i$ assuming the same
frequencies.

\section{Applications}

\subsection{Spherical nanocrystals with cubic crystallinity}

\subsubsection{Spherical mono-domain silver and gold nanoparticles}

\begin{figure}[!ht]
 \begin{center}
  \includegraphics[width=\columnwidth]{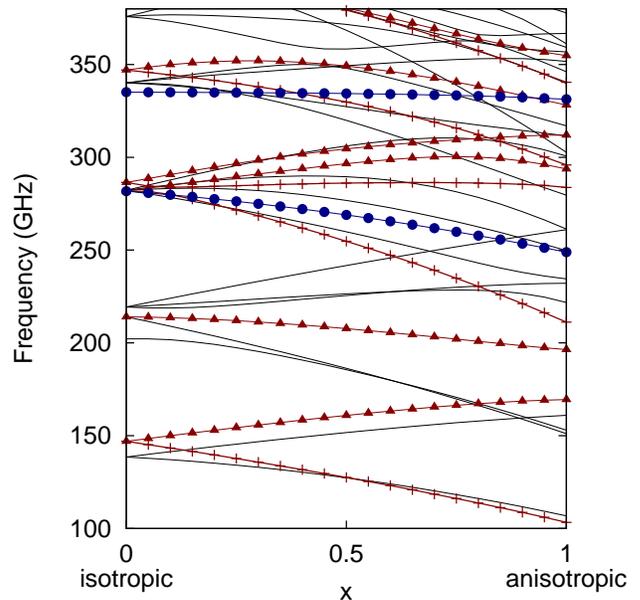}
 \end{center}
 \caption{\label{FigAg}(Color online) Evolution of the frequency
of Raman active modes of a mono-domain
 silver sphere (radius 5~nm) with varying
elastic anisotropy up to the frequency range of the breathing mode.
Raman inactive modes are plotted with black lines,
E$_{\text{g}}$ branches with lines and crosses (red online), T$_{\text{2g}}$
branches with lines and triangles (red online) and A$_{\text{1g}}$
branches with lines and full circles (blue online).}
\end{figure}

Let us first consider the case of a spherical nanoparticle made
of silver.  This nanoparticle is mono-domain and therefore the
stiffness tensor is the same everywhere inside the nanoparticle
and identical to that of bulk silver.\cite{NeighboursPR58}
Table~\ref{TabAg} presents the calculated frequencies,
irreducible representations, main projections onto the modes of
an isotropic silver sphere and volume variations of the lowest
frequency modes. The six modes $i = 1 \ldots 6$ have zero
frequency and correspond to the rigid rotations and translations
of the nanoparticle. The branches corresponding to the lowering
of the symmetry when going from the isotropic to the anisotropic
case are plotted in figure~\ref{FigAg}.

\begin{table}
\begin{tabular}{c|c|c|c|c}
  $i$ & $\nu$ (GHz) & i.r. & Squared Lamb projections & $\delta V$\\
\hline
7-8 & 103.3 
 & E$_{\text{g}}$
 & 0.995 S$_2^1$ + 0.002 S$_4^1$ +  \ldots
 & 0.0 \\
9-11 & 106.8 
 & T$_{\text{2u}}$
 & 0.936 T$_2^1$ + 0.054 S$_3^1$ +  \ldots
 & 0.0 \\
12 & 151.1
 & A$_{\text{2g}}$
 & 0.996 T$_3^1$ + 0.002 S$_6^1$ +  \ldots
 & 0.0 \\
13-15 & 153.0 
 & T$_{\text{1u}}$
 & 0.499 S$_3^1$ + 0.480 S$_1^1$ +  \ldots
 & 0.0 \\
16-17 & 161.0 
 & E$_{\text{u}}$
 & 0.974 T$_2^1$ + 0.018 T$_4^1$ +  \ldots
 & 0.0 \\
18-20 & 169.4 
 & T$_{\text{2g}}$
 & 0.821 S$_2^1$ + 0.160 T$_3^1$ +  \ldots
 & 0.0 \\
21-23 & 196.4 
 & T$_{\text{2g}}$
 & 0.758 T$_3^1$ + 0.178 S$_2^1$ +  \ldots
 & 0.0 \\
24-26 & 196.9 
 & T$_{\text{1g}}$
 & 0.411 T$_3^1$ + 0.369 S$_4^1$ +  \ldots
 & 0.0 \\
27-28 & 211.2 
 & E$_{\text{g}}$
 & 0.838 S$_2^2$ + 0.113 S$_4^1$ +  \ldots
 & 0.0 \\
29-31 & 221.8 
 & T$_{\text{2u}}$
 & 0.748 T$_4^1$ + 0.199 S$_3^1$ +  \ldots
 & 0.0 \\
32-34 & 232.1 
 & T$_{\text{1u}}$
 & 0.466 S$_3^1$ + 0.466 S$_1^1$ +  \ldots
 & 0.0 \\
35-37 & 234.5 
 & T$_{\text{2u}}$
 & 0.665 S$_3^1$ + 0.127 T$_4^1$ +  \ldots
 & 0.0 \\
38 & 248.9
 & A$_{\text{1g}}$
 & 0.959 S$_4^1$ + 0.028 S$_0^1$ +  \ldots
 & 0.4 \\
\ldots & \ldots & \ldots & \ldots & \ldots \\
75 & 331.2
 & A$_{\text{1g}}$
 & 0.930 S$_0^1$ + 0.039 S$_4^2$ +  \ldots
 & 4.3 \\
\ldots & \ldots & \ldots & \ldots & \ldots \\
\end{tabular}
\caption{\label{TabAg}
Characteristics of the modes of a mono-domain
Ag sphere with $R=5$~nm are shown. $i$ is the mode
index. The two largest projections onto Lamb modes
are shown.}
\end{table}

Figure~\ref{FigAg} clearly shows that the introduction of elastic
anisotropy significantly lifts the degeneracy of
most modes. One notable exception is the breathing mode which
corresponds to the second
A$_{\text{1g}}$ branch. This mode is non-degenerate and its frequency
hardly changes with anisotropy.
The other important exception are the dipolar modes S$_1$ which are
infrared active and transform into T$_{\text{1u}}$ with the
same degeneracy.
The lowest Raman active mode S$_2^1$
which has degeneracy 5 is split into the lowest E$_{\text{g}}$ and
T$_{\text{2g}}$ branches ($i=7,8$ and $i=18,19,20$ in table~\ref{TabAg}
respectively) as expected from table~\ref{degeneracy}. These
anisotropic modes have a dominant projection onto S$_2^1$ confirming
their Lamb mode parentage.

\begin{figure}[!ht]
 \begin{center}
  \includegraphics[width=\columnwidth]{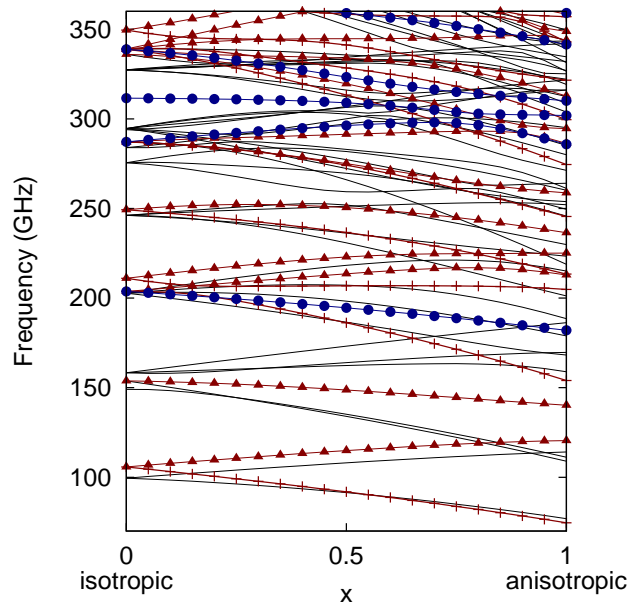}
 \end{center}
 \caption{\label{FigAu}
(Color online) Evolution of the frequency of Raman active modes
of a mono-domain gold sphere (radius 5~nm) with varying elastic
anisotropy up to the frequency range of the breathing mode.
Raman inactive modes are plotted with black lines,
E$_{\text{g}}$ branches with lines and crosses (red online), T$_{\text{2g}}$
branches with lines and triangles (red online) and A$_{\text{1g}}$
branches with lines and full circles (blue online).}
\end{figure}

Similar calculations have been performed for mono-domain gold
nanoparticles using the elastic constants from Ref.~\onlinecite{LBAu}
and the results are presented in table~\ref{TabAu} and
figure~\ref{FigAu}. Compared to the previous case of silver, only the
values of the stiffness tensor were changed so most observations
previously made still apply.
We recently reported on the experimental observation of the
splitting of S$_2^1$ for such nanoparticles.\cite{PortalesPNAS08}
There is an excellent agreement between these measurements and
the splitting calculated using the present approach.  This
strongly supports the validity of our approach and in particular
the relevance of elastic anisotropy even in such very small
nanoparticles.

\begin{figure}[!ht]
 \begin{center}
  \includegraphics[width=\columnwidth]{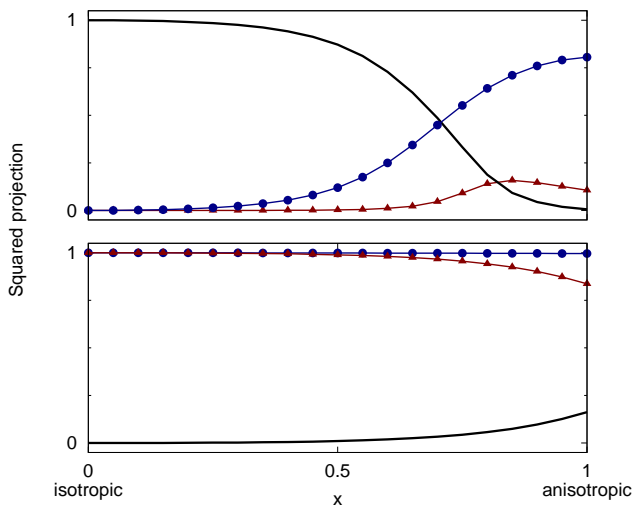}
 \end{center}
 \caption{\label{FigProjAu}(Color online) Evolution of the squared
projection of several modes of a mono-domain gold sphere (radius 5~nm)
with varying elastic anisotropy. The bottom plot shows the squared
projections
onto S$_2^1$ of the E$_{\text{g}}$ branch coming from S$_2^1$ (line
with circles, blue online) and the T$_{\text{2g}}$ branches coming from
S$_2^1$ (line with triangles, red online) and T$_3^1$ (black line).
The upper plot shows the squared projections onto S$_0^1$ of the
A$_{\text{1g}}$ branches coming from S$_6^1$ (line and triangles, red
online), S$_0^1$ (black line) and S$_4^2$ (line and circles,
blue online). }
\end{figure}

One notable difference between gold and silver concerns the case
of the breathing mode.  The isotropic breathing mode at 311.6~GHz
belongs to the third A$_{\text{1g}}$ branch.  Therefore it is
tempting to assume that the anisotropic breathing mode lies on
the same branch and is therefore mode $i=133$ at 301.9~GHz.
However, table~\ref{TabAu} reveals that mode $i=141$ is a much
better candidate for a breathing mode due to its larger volume
variation and projection onto S$_0^1$. It turns out this is due
to a strong mixing of the A$_{\text{1g}}$ branches as anisotropy
is increased. Indeed, the isotropic modes S$_4$, S$_6$ and S$_8$
are split into various irreducible representations, one of them
being A$_{\text{1g}}$. Unlike the previous case of silver, the
elastic constants of gold result in four A$_{\text{1g}}$
branches coming from S$_0^1$, S$_4^2$, S$_6^1$ and S$_8^1$ being in
the same frequency range. As a result, mode $i=141$ which is the mode
having the strongest projection onto S$_0^1$ and also has a large
volume variation does not lie on the same branch as S$_0^1$.
Figure~\ref{FigProjAu} shows the variation of the squared projections onto
S$_0^1$ of three of these A$_{\text{1g}}$ branches as the elastic anisotropy is
varied and confirm the mixing discussed above.

\begin{table}
\begin{tabular}{c|c|c|c|c}
 $i$  & $\nu$ (GHz) & i.r. & Squared Lamb projections & $\delta V$ \\
\hline
7-8 & 74.6 
 & E$_{\text{g}}$
 & 0.996 S$_2^1$ + 0.002 S$_4^1$ +  \ldots
 & 0.0 \\
9-11 & 76.9 
 & T$_{\text{2u}}$
 & 0.939 T$_2^1$ + 0.053 S$_3^1$ +  \ldots
 & 0.0 \\
12 & 109.0
 & A$_{\text{2g}}$
 & 0.996 T$_3^1$ + 0.002 S$_6^1$ +  \ldots
 & 0.0 \\
13-15 & 111.2 
 & T$_{\text{1u}}$
 & 0.516 S$_3^1$ + 0.466 S$_1^1$ +  \ldots
 & 0.0 \\
16-17 & 114.1 
 & E$_{\text{u}}$
 & 0.977 T$_2^1$ + 0.016 T$_4^1$ +  \ldots
 & 0.0 \\
18-20 & 120.5 
 & T$_{\text{2g}}$
 & 0.837 S$_2^1$ + 0.146 T$_3^1$ +  \ldots
 & 0.0 \\
21-23 & 140.2 
 & T$_{\text{2g}}$
 & 0.773 T$_3^1$ + 0.162 S$_2^1$ +  \ldots
 & 0.0 \\
24-26 & 141.8 
 & T$_{\text{1g}}$
 & 0.433 T$_3^1$ + 0.356 S$_4^1$ +  \ldots
 & 0.0 \\
27-28 & 154.0 
 & E$_{\text{g}}$
 & 0.824 S$_2^2$ + 0.126 S$_4^1$ +  \ldots
 & 0.0 \\
29-31 & 158.9 
 & T$_{\text{2u}}$
 & 0.698 T$_4^1$ + 0.259 S$_3^1$ +  \ldots
 & 0.0 \\
32-34 & 168.2 
 & T$_{\text{2u}}$
 & 0.617 S$_3^1$ + 0.186 T$_4^1$ +  \ldots
 & 0.0 \\
35-37 & 169.7 
 & T$_{\text{1u}}$
 & 0.484 S$_1^1$ + 0.456 S$_3^1$ +  \ldots
 & 0.0 \\
\ldots & \ldots & \ldots & \ldots & \ldots \\
43 & 182.0
 & A$_{\text{1g}}$
 & 0.987 S$_4^1$ + 0.005 S$_0^1$ +  \ldots
 & 0.1 \\
\ldots & \ldots & \ldots & \ldots & \ldots \\
118 & 285.8
 & A$_{\text{1g}}$
 & 0.594 S$_4^2$ + 0.218 S$_8^1$ +  \ldots
 & 1.2 \\
\ldots & \ldots & \ldots & \ldots & \ldots \\
133 & 301.9
 & A$_{\text{1g}}$
 & 0.866 S$_6^1$ + 0.049 S$_8^1$ +  \ldots
 & 0.3 \\
\ldots & \ldots & \ldots & \ldots & \ldots \\
141 & 310.1
 & A$_{\text{1g}}$
 & 0.806 S$_0^1$ + 0.133 S$_8^1$ +  \ldots
 & 3.9 \\
\ldots & \ldots & \ldots & \ldots & \ldots \\
183 & 341.5
 & A$_{\text{1g}}$
 & 0.378 S$_8^1$ + 0.319 S$_4^2$ +  \ldots
 & 1.4 \\
\ldots & \ldots & \ldots & \ldots & \ldots \\
\end{tabular}
\caption{\label{TabAu}
Characteristics of the modes of a mono-domain
sphere of Au with $R=5$~nm. $i$ is the mode index.
The two largest projections onto the Lamb modes
are shown.}
\end{table}

Similar but less pronounced mixings are observed in all cases.
The density of modes (\textit{i.e.} the number of modes per unit
frequency) increases with frequency. As a result, the probability
of mixings increases with frequency or $i$.  This explains why
the lowest frequency modes such as the lowest E$_{\text{g}}$ ones
are almost pure isotropic modes (S$_2^1$).  However, the lowest
T$_{\text{2g}}$ branches which are issued from the same S$_2^1$
and from T$_3^1$ are already significantly mixed by anisotropy
for both gold and silver nanoparticles.  This is clearly
evidenced in figure~\ref{FigProjAu} too.

\subsubsection{Raman scattering efficiency for metallic cubic materials}

Low-frequency Raman scattering from silver and gold nanoparticles
has attracted a lot of attention during the last few decades. The scattering
mechanisms have been identified\cite{BachelierPRB04} and one might
wonder how the anisotropic considerations detailed in this work fit into
this picture. The inelastic light scattering process for such
nanoparticles is mediated by the dipolar plasmon. As a result only
scattering by S$_0$ and S$_2$ modes is allowed\cite{duval92}
and has a significant Raman scattering
cross-section\cite{BachelierPRB04} in the isotropic case.
Anisotropic nanoparticles obey the same rules and one might
qualitatively estimate their Raman intensity by
using the projections onto the same isotropic modes.
As a result
a significant scattering intensity is expected for the lowest
E$_{\text{g}}$ and T$_{\text{2g}}$ modes of both silver and gold while
the second T$_{\text{2g}}$ mode should have a small Raman cross-section
due to its small projection onto  S$_2^1$. The volume mechanism enables
a measurable Raman intensity for the A$_{\text{1g}}$ modes having a
significant projection onto S$_0$, \textit{i.e.} for modes $i=75$ and
$i=141$ for silver and gold respectively. The Raman intensity for other
A$_{\text{1g}}$ modes should be very weak and probably not measurable in
practice.

\subsubsection{Other cubic materials} 

For cubic materials, the degree of elastic anisotropy
is quantified by the Zener anisotropy ratio,
$A$ = $2 C_{44} / (C_{11} - C_{12})$. $A$=$1$ for an isotropic
material. Silver and copper both have $A \simeq 3$ and also
had the largest mode splittings that we found. For comparison,
we mention results for some less anisotropic materials.
For a silicon nanosphere with radius 5~nm, the
Lamb mode
S$_2^1$ splits into
two frequencies at 386.6~GHz (E$_{\text{g}}$) and 485.9~GHz
(T$_{\text{2g}}$). For a germanium nanosphere, the same mode is split
into 235.1~GHz (E$_{\text{g}}$) and 295.3~GHz (T$_{\text{2g}}$).
Using
the standard deviation as a simple measure of the frequency splitting
$\Delta$ (and therefore neglecting the effect of mixings), we obtain
$\Delta = $23, 22, 10 and 11\% for Ag, Au, Si and Ge respectively. For
all these commonly studied materials, this splitting cannot be neglected.

\subsection{Spherical nanocrystals with tetragonal crystallinity}

\begin{figure}[!ht]
 \begin{center}
  \includegraphics[width=\columnwidth]{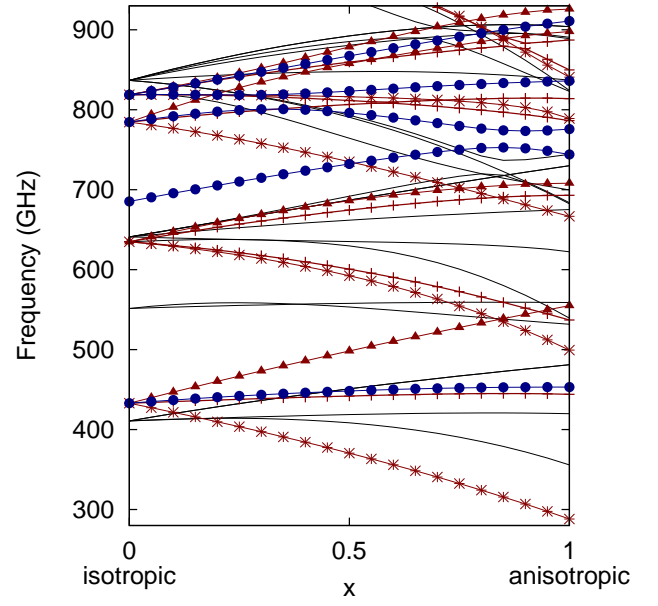}
 \end{center}
 \caption{\label{FigTiO2r}(Color online) Evolution of the frequency
of Raman active modes of a mono-domain
rutile TiO$_2$ sphere (radius 5~nm) with varying
elastic anisotropy up to the frequency range of the breathing mode.
Raman inactive modes are plotted with black lines,
E$_{\text{g}}$ branches with lines and crosses (red online), B$_{\text{1g}}$
branches with lines and asterisks (red online), B$_{\text{2g}}$
branches with lines and triangles (red online) and A$_{\text{1g}}$
branches with lines and full circles (blue online).}
\end{figure}

Nanospheres with tetragonal crystallinity have a lower symmetry than
the previous nanospheres with cubic crystallinity. The corresponding
point group is D$_{\text{4h}}$. As a result, more degeneracy lifting
occurs. This results in a splitting of the infrared active S$_1$ mode
into A$_{\text{2u}}$ and E$_{\text{u}}$ and 
four branches starting from the S$_2^1$ mode (A$_{\text{1g}}$,
B$_{\text{1g}}$, B$_{\text{2g}}$ and E$_{\text{g}}$). The
relative order
of these branches depends on the stiffness
tensor. Two nanospheres made of TiO$_2$ with radius 5~nm will be
considered using
the parameters from Ref.~\onlinecite{IugaEPJ07}.
One of them has the anatase crystal structure
(table~\ref{TabTiO2a})
and the other the rutile structure 
(table~\ref{TabTiO2r} and figure~\ref{FigTiO2r}).
We used these calculations for the anatase crystal structure in a recent
work\cite{SaviotPRB08} to model the inelastic scattering of neutrons.
Both have tetragonal symmetry but
different elastic parameters. This results in different relative
positions of the frequencies of the modes coming from a given isotropic
mode.

Using the same measure as before, the frequency splitting for the
S$_2^1$ modes is $\Delta =$~10 and 20\% for the anatase and
rutile structure respectively.

Due to the lowering of the symmetry compared to the cubic case,
the breathing mode is also susceptible to more mixings.  For
anatase TiO$_2$, the S$_0^1$ mode mixes mainly with T$_5^1$ (mode
$i=67$) while for rutile TiO$_2$ it mixes mainly with S$_2^2$
(mode $i=41$).

\begin{table}
\begin{tabular}{c|c|c|c|c}
  $i$  & $\nu$ (GHz) & i.r. & Squared Lamb projections & $\delta V$ \\
\hline
7 & 283.5
 & A$_{\text{1g}}$
 & 0.977 S$_2^1$ + 0.016 S$_0^1$ +  \ldots
 & 0.6 \\
8-9 & 284.0 
 & E$_{\text{u}}$
 & 0.931 T$_2^1$ + 0.054 S$_1^1$ +  \ldots
 & 0.0 \\
10 & 298.5
 & A$_{\text{1u}}$
 & 0.999 T$_2^1$ + 0.000 T$_9^1$ +  \ldots
 & 0.0 \\
11 & 304.3
 & B$_{\text{1u}}$
 & 0.998 T$_2^1$ + 0.001 S$_3^1$ +  \ldots
 & 0.0 \\
12-13 & 316.1 
 & E$_{\text{g}}$
 & 0.996 S$_2^1$ + 0.002 T$_3^1$ +  \ldots
 & 0.0 \\
14 & 326.9
 & B$_{\text{2g}}$
 & 0.992 S$_2^1$ + 0.008 T$_3^1$ +  \ldots
 & 0.0 \\
15 & 330.9
 & B$_{\text{2u}}$
 & 0.967 T$_2^1$ + 0.027 S$_3^1$ +  \ldots
 & 0.0 \\
16 & 381.5
 & B$_{\text{1g}}$
 & 0.965 S$_2^1$ + 0.033 T$_3^1$ +  \ldots
 & 0.0 \\
17 & 432.4
 & A$_{\text{2u}}$
 & 0.494 S$_1^1$ + 0.436 S$_3^1$ +  \ldots
 & 0.0 \\
18 & 443.3
 & B$_{\text{2g}}$
 & 0.950 T$_3^1$ + 0.028 S$_2^2$ +  \ldots
 & 0.0 \\
19-20 & 445.7 
 & E$_{\text{u}}$
 & 0.612 S$_1^1$ + 0.323 S$_3^1$ +  \ldots
 & 0.0 \\
21 & 447.6
 & A$_{\text{2u}}$
 & 0.537 S$_3^1$ + 0.448 S$_1^1$ +  \ldots
 & 0.0 \\
22-23 & 455.5 
 & E$_{\text{g}}$
 & 0.957 T$_3^1$ + 0.015 S$_4^1$ +  \ldots
 & 0.0 \\
24-25 & 463.4 
 & E$_{\text{u}}$
 & 0.758 S$_3^1$ + 0.220 S$_1^1$ +  \ldots
 & 0.0 \\
26 & 470.7
 & A$_{\text{2g}}$
 & 0.989 T$_3^1$ + 0.005 T$_1^1$ +  \ldots
 & 0.0 \\
27 & 473.9
 & B$_{\text{1g}}$
 & 0.865 T$_3^1$ + 0.060 S$_4^1$ +  \ldots
 & 0.0 \\
28 & 476.9
 & B$_{\text{1u}}$
 & 0.987 S$_3^1$ + 0.008 T$_4^1$ +  \ldots
 & 0.0 \\
29-30 & 480.7 
 & E$_{\text{g}}$
 & 0.941 T$_3^1$ + 0.031 S$_4^1$ +  \ldots
 & 0.0 \\
31 & 510.9
 & B$_{\text{2u}}$
 & 0.946 S$_3^1$ + 0.026 T$_4^1$ +  \ldots
 & 0.0 \\
32-33 & 537.1 
 & E$_{\text{u}}$
 & 0.826 S$_3^1$ + 0.095 S$_1^1$ +  \ldots
 & 0.0 \\
34 & 573.8
 & A$_{\text{1g}}$
 & 0.974 S$_4^1$ + 0.009 S$_4^2$ +  \ldots
 & 0.0 \\
35 & 580.8
 & A$_{\text{1g}}$
 & 0.910 S$_2^2$ + 0.030 S$_2^3$ +  \ldots
 & 0.8 \\
\ldots & \ldots & \ldots & \ldots & \ldots \\
55 & 678.4
 & A$_{\text{1g}}$
 & 0.885 S$_4^1$ + 0.082 T$_5^1$ +  \ldots
 & 0.2 \\
\ldots & \ldots & \ldots & \ldots & \ldots \\
67 & 743.1
 & A$_{\text{1g}}$
 & 0.803 S$_0^1$ + 0.156 T$_5^1$ +  \ldots
 & 4.0 \\
\ldots & \ldots & \ldots & \ldots & \ldots \\
69 & 751.6
 & A$_{\text{1g}}$
 & 0.675 T$_5^1$ + 0.162 S$_0^1$ +  \ldots
 & 1.8 \\
\ldots & \ldots & \ldots & \ldots & \ldots \\
\end{tabular}
\caption{\label{TabTiO2a}
Characteristics of the modes of a mono-domain anatase TiO$_2$
sphere with $R=5$~nm are shown. $i$ is the mode index. The two
largest projections onto Lamb modes are shown.}
\end{table}


\begin{table}
\begin{tabular}{c|c|c|c|c}
  $i$  & $\nu$ (GHz) & i.r. & Squared Lamb projections & $\delta V$ \\
\hline
7 & 288.1
 & B$_{\text{1g}}$
 & 0.978 S$_2^1$ + 0.013 T$_3^1$ +  \ldots
 & 0.0 \\
8 & 296.9
 & B$_{\text{2u}}$
 & 0.943 T$_2^1$ + 0.045 S$_3^1$ +  \ldots
 & 0.0 \\
9-10 & 355.8 
 & E$_{\text{u}}$
 & 0.693 T$_2^1$ + 0.150 S$_3^1$ +  \ldots
 & 0.0 \\
11 & 419.8
 & A$_{\text{1u}}$
 & 0.981 T$_2^1$ + 0.013 T$_4^1$ +  \ldots
 & 0.0 \\
12-13 & 444.1 
 & E$_{\text{g}}$
 & 0.927 S$_2^1$ + 0.062 T$_3^1$ +  \ldots
 & 0.0 \\
14 & 453.1
 & A$_{\text{1g}}$
 & 0.955 S$_2^1$ + 0.035 S$_0^1$ +  \ldots
 & 0.8 \\
15 & 481.0
 & B$_{\text{1u}}$
 & 0.965 T$_2^1$ + 0.033 S$_3^1$ +  \ldots
 & 0.0 \\
16 & 499.4
 & B$_{\text{1g}}$
 & 0.715 T$_3^1$ + 0.210 S$_2^2$ +  \ldots
 & 0.0 \\
17-18 & 531.7 
 & E$_{\text{u}}$
 & 0.402 S$_1^1$ + 0.288 T$_2^1$ +  \ldots
 & 0.0 \\
19-20 & 537.1 
 & E$_{\text{g}}$
 & 0.813 T$_3^1$ + 0.071 S$_2^1$ +  \ldots
 & 0.0 \\
21 & 539.4
 & A$_{\text{2g}}$
 & 0.465 T$_3^1$ + 0.327 S$_4^1$ +  \ldots
 & 0.0 \\
22 & 555.0
 & B$_{\text{2g}}$
 & 0.997 S$_2^1$ + 0.001 S$_4^1$ +  \ldots
 & 0.0 \\
23 & 558.9
 & A$_{\text{2u}}$
 & 0.969 S$_1^1$ + 0.016 S$_3^1$ +  \ldots
 & 0.0 \\
24 & 600.0
 & B$_{\text{2u}}$
 & 0.845 S$_3^1$ + 0.095 T$_4^1$ +  \ldots
 & 0.0 \\
25-26 & 622.3 
 & E$_{\text{u}}$
 & 0.575 S$_3^1$ + 0.247 S$_1^1$ +  \ldots
 & 0.0 \\
\ldots & \ldots & \ldots & \ldots & \ldots \\
38 & 744.0
 & A$_{\text{1g}}$
 & 0.739 S$_4^1$ + 0.157 S$_0^1$ +  \ldots
 & 1.4 \\
39-40 & 744.2 
 & E$_{\text{u}}$
 & 0.725 S$_3^1$ + 0.178 S$_1^1$ +  \ldots
 & 0.0 \\
41 & 775.8
 & A$_{\text{1g}}$
 & 0.631 S$_0^1$ + 0.269 S$_2^2$ +  \ldots
 & 4.1 \\
\ldots & \ldots & \ldots & \ldots & \ldots \\
50 & 836.1
 & A$_{\text{1g}}$
 & 0.439 S$_2^2$ + 0.351 S$_4^1$ +  \ldots
 & 1.6 \\
\ldots & \ldots & \ldots & \ldots & \ldots \\
62 & 911.0
 & A$_{\text{1g}}$
 & 0.739 S$_4^1$ + 0.169 S$_2^2$ +  \ldots
 & 0.9 \\
\ldots & \ldots & \ldots & \ldots & \ldots \\
81 & 1019.1
 & A$_{\text{1g}}$
 & 0.649 T$_5^1$ + 0.110 S$_6^1$ +  \ldots
 & 0.5 \\
\ldots & \ldots & \ldots & \ldots & \ldots \\
\end{tabular}
\caption{\label{TabTiO2r}
Characteristics of the modes of a mono-domain rutile TiO$_2$
sphere with $R=5$~nm are shown. $i$ is the mode index. The two
largest projections onto Lamb modes are shown.}
\end{table}

\subsection{Spherical nanocrystals with hexagonal crystallinity}

\begin{figure}[!ht]
 \begin{center}
  \includegraphics[width=\columnwidth]{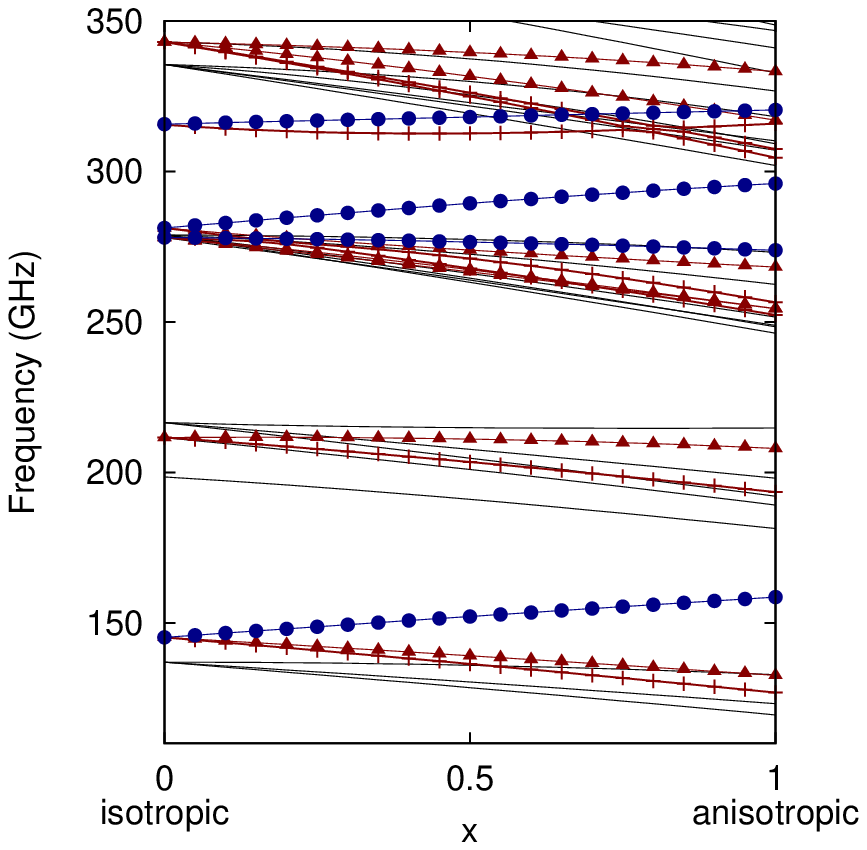}
 \end{center}
 \caption{\label{FigCdSe}(Color online) Evolution of the frequency
of Raman active modes of a mono-domain wurtzite CdSe sphere (radius
5~nm) with varying elastic anisotropy up to the frequency range of the
breathing mode. Raman inactive modes are plotted with black lines,
E$_{\text{1g}}$ branches with lines and crosses (red online),
E$_{\text{2g}}$ branches with lines and triangles (red online) and
A$_{\text{1g}}$ branches with lines and full circles (blue online).}
\end{figure}

As a last example of the influence of elastic anisotropy, we
focus on nanospheres made of crystals having hexagonal
symmetry, namely CdSe
(table~\ref{TabCdSe} and figure~\ref{FigCdSe} using the elastic
constants from Ref.~\onlinecite{RabaniJCP02}),
Co (table~\ref{TabCo} using the elastic
constants from Ref.~\onlinecite{GangopadhyayAPL07}) and ZnO
(elastic constants from Ref.~\onlinecite{CombePRB09}).
The associated point group is D$_{\text{6h}}$.
The most notable difference compared to previous symmetries is that some
``accidental'' degeneracy exists for such systems: for each
B$_{\text{1g}}$ (B$_{\text{1u}}$) vibration there is a B$_{\text{2g}}$
(B$_{\text{2u}}$) vibration having the same frequency.

Using the same measure as before, the frequency splitting for the
S$_2^1$ modes is $\Delta =$9, 11 and 5\% for the CdSe, Co and ZnO
respectively. ZnO is therefore the most elastically isotropic amongst
the materials studied in this work.

Regarding the breathing mode, for CdSe it is hardly mixed with
other modes due to anisotropy and results in mode $i = 77$ which
has a strong projection onto S$_0^1$ and a large volume
variation. For the nanosphere made of cobalt, there is a strong
mixing with the S$_2^2$ and S$_4^1$ modes and the anisotropic
mode with the largest projection onto S$_0^1$ and the largest
volume variation is mode $i = 49$ which is on the branch coming
from S$_4^1$ at 466.7~GHz and not on the branch starting from
S$_0^1$ at 536.8~GHz.

The infrared active S$_1^1$ mode which was recently
observed\cite{LiuAPL08} for CdSe nanoparticles is split by the
lowering of symmetry into A$_{\text{2u}}$ and E$_{\text{1u}}$
modes at 181.9 and 181.3~GHz respectively. This frequency
splitting is very small but these modes mix with the branches
having identical irreducible representations coming from S$_3^1$.
Neglecting this mixing, the isotropic S$_1^1$ is in excellent
agreement with the anisotropic description.

\begin{table}
\begin{tabular}{c|c|c|c|c}
  $i$  & $\nu$ (GHz) & i.r. & Squared Lamb projections & $\delta V$ \\
\hline
7 & 119.4
 & A$_{\text{1u}}$
 & 1.000 T$_2^1$ + 0.000 T$_8^1$ +  \ldots 
 & 0.0 \\
8-9 & 123.2 
 & E$_{\text{2u}}$
 & 0.998 T$_2^1$ + 0.001 S$_3^1$ +  \ldots 
 & 0.0 \\
10-11 & 126.8 
 & E$_{\text{1g}}$
 & 0.999 S$_2^1$ + 0.001 T$_3^1$ +  \ldots 
 & 0.0 \\
12-13 & 132.6 
 & E$_{\text{2g}}$
 & 0.998 S$_2^1$ + 0.002 T$_3^1$ +  \ldots 
 & 0.0 \\
14-15 & 132.8 
 & E$_{\text{1u}}$
 & 0.971 T$_2^1$ + 0.023 S$_3^1$ +  \ldots 
 & 0.0 \\
16 & 158.6
 & A$_{\text{1g}}$
 & 0.999 S$_2^1$ + 0.001 S$_4^1$ +  \ldots 
 & 0.0 \\
17-18 & 181.3 
 & E$_{\text{1u}}$
 & 0.894 S$_1^1$ + 0.103 S$_3^1$ +  \ldots 
 & 0.0 \\
19 & 181.9
 & A$_{\text{2u}}$
 & 0.789 S$_1^1$ + 0.206 S$_3^1$ +  \ldots 
 & 0.0 \\
20 & 186.1
 & A$_{\text{2g}}$
 & 0.999 T$_3^1$ + 0.001 T$_1^1$ +  \ldots 
 & 0.0 \\
21-22 & 189.2 
 & B$_{\text{1g}}$+B$_{\text{2g}}$
 & 0.994 T$_3^1$ + 0.004 S$_4^1$ +  \ldots 
 & 0.0 \\
23-24 & 192.1 
 & E$_{\text{2u}}$
 & 0.998 S$_3^1$ + 0.001 T$_2^1$ +  \ldots 
 & 0.0 \\
25-26 & 193.5 
 & E$_{\text{1g}}$
 & 0.980 T$_3^1$ + 0.014 S$_4^1$ +  \ldots 
 & 0.0 \\
27-28 & 198.1 
 & B$_{\text{1u}}$+B$_{\text{2u}}$
 & 0.996 S$_3^1$ + 0.004 T$_4^1$ +  \ldots 
 & 0.0 \\
29-30 & 208.0 
 & E$_{\text{2g}}$
 & 0.947 T$_3^1$ + 0.042 S$_4^1$ +  \ldots 
 & 0.0 \\
31-32 & 214.7 
 & E$_{\text{1u}}$
 & 0.870 S$_3^1$ + 0.099 S$_1^1$ +  \ldots 
 & 0.0 \\
33 & 226.7
 & A$_{\text{2u}}$
 & 0.789 S$_3^1$ + 0.206 S$_1^1$ +  \ldots 
 & 0.0 \\
\ldots & \ldots & \ldots & \ldots & \ldots \\
55 & 273.8
 & A$_{\text{1g}}$
 & 0.926 S$_4^1$ + 0.060 S$_2^2$ +  \ldots 
 & 0.3 \\
56 & 284.9
 & A$_{\text{2g}}$
 & 0.999 T$_1^1$ + 0.001 T$_3^1$ +  \ldots 
 & 0.0 \\
57 & 296.0
 & A$_{\text{1g}}$
 & 0.925 S$_2^2$ + 0.059 S$_4^1$ +  \ldots 
 & 0.2 \\
\ldots & \ldots & \ldots & \ldots & \ldots \\
77 & 320.4
 & A$_{\text{1g}}$
 & 0.985 S$_0^1$ + 0.011 S$_4^1$ +  \ldots 
 & 4.5 \\
\ldots & \ldots & \ldots & \ldots & \ldots \\
\end{tabular}
\caption{\label{TabCdSe}
Characteristics of the modes of a mono-domain wurtzite CdSe
sphere with $R=5$~nm are shown. $i$ is the mode index. The two
largest projections onto Lamb modes are shown.}
\end{table}

\begin{table}
\begin{tabular}{c|c|c|c|c}
  $i$  & $\nu$ (GHz) & i.r. & Squared Lamb projections & $\delta V$ \\
\hline
7-8 & 229.8 
 & E$_{\text{2u}}$
 & 0.998 T$_2^1$ + 0.001 T$_4^1$ +  \ldots 
 & 0.0 \\
9 & 232.2
 & A$_{\text{1u}}$
 & 1.000 T$_2^1$ + 0.000 T$_8^1$ +  \ldots 
 & 0.0 \\
10-11 & 240.7 
 & E$_{\text{2g}}$
 & 0.997 S$_2^1$ + 0.003 T$_3^1$ +  \ldots 
 & 0.0 \\
12-13 & 246.6 
 & E$_{\text{1g}}$
 & 1.000 S$_2^1$ + 0.000 T$_8^1$ +  \ldots 
 & 0.0 \\
14-15 & 251.4 
 & E$_{\text{1u}}$
 & 0.942 T$_2^1$ + 0.029 S$_1^1$ +  \ldots 
 & 0.0 \\
16 & 316.1
 & A$_{\text{1g}}$
 & 0.996 S$_2^1$ + 0.003 S$_2^2$ +  \ldots 
 & 0.0 \\
17-18 & 334.7 
 & E$_{\text{1u}}$
 & 0.919 S$_1^1$ + 0.063 S$_3^1$ +  \ldots 
 & 0.0 \\
19 & 346.4
 & A$_{\text{2u}}$
 & 0.828 S$_1^1$ + 0.167 S$_3^1$ +  \ldots 
 & 0.0 \\
\ldots & \ldots & \ldots & \ldots & \ldots \\
49 & 497.8
 & A$_{\text{1g}}$
 & 0.958 S$_0^1$ + 0.042 S$_4^1$ +  \ldots 
 & 4.5 \\
\ldots & \ldots & \ldots & \ldots & \ldots \\
\end{tabular}
\caption{\label{TabCo}
Characteristics of the modes of a mono-domain Co sphere with
$R=5$~nm are shown. $i$ is the mode index. The two largest
projections onto Lamb modes are shown.}
\end{table}


\subsection{Non-spherical nanocrystals}

In order to illustrate the usefulness of the same numerical tools
for a different source of anisotropy, we consider in the following the
lowering of the symmetry due to the shape of the nanoparticles. We start
with a minor change of the shape as the nanosphere is transformed into a
spheroid having an isotropic elasticity. Then we consider faceted
nanoparticles with elastic anisotropy.

\subsubsection{Spheroids with isotropic elasticity}

\begin{figure}[!ht]
 \begin{center}
  \includegraphics[width=\columnwidth]{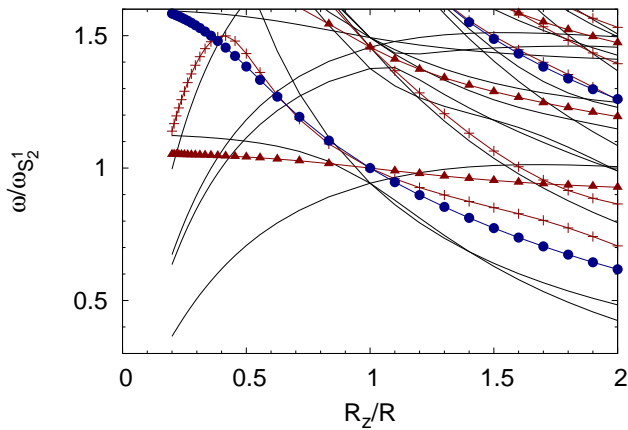}
 \end{center}
 \caption{\label{FigSpheroid}Reduced frequency of the vibrations
of an elastically isotropic silver spheroid as a function of its
aspect ratio. $R$ is the length of the degenerate semi-axis and $R_z$
is the length of the non-degenerate one.
$\omega_{\text{S}_2^1}$
is the frequency of the S$_2^1$ mode of the sphere of radius
$R$. Raman inactive modes are plotted with black lines,
E$_{\text{1g}}$ branches with lines and crosses (red online),
E$_{\text{2g}}$ branches with lines and triangles (red online) and
A$_{\text{1g}}$ branches with lines and full circles (blue online).}
\end{figure}

Let us consider a spheroid made of silver with a degenerate
semi-axis $R = 5$~nm and a varying non-degenerate semi-axis $R_z$.
We assume the
elasticity to be isotropic. The point group associated with
such a
spheroid is therefore D$_{\infty \text{h}}$. The frequencies,
irreducible representations and volume variations for a spheroid with
$R_z = 10$~nm are presented in table~\ref{TabSpheroid}. The branches
obtained for varying $R_z$ are presented in figure~\ref{FigSpheroid}.

Let us now focus on the effect of the spheroidal deformation on
the S$_2^1$ modes. Looking at the displacements corresponding to
these modes for small deviations from the sphere, it is possible
to understand the frequency variations.  The lowest
A$_{\text{1g}}$ mode corresponds to a stretching along the $z$
direction accompanied by a shrinking in the $xy$ plane. Therefore
it can be seen as a vibration confined along the $z$ direction
and its frequency varies roughly as $1/R_z$.  The E$_{\text{2g}}$
vibrations correspond to a stretching in the $xy$ plane without
changes along the $z$ axis and therefore their frequencies hardly
changes with $R_z$.  The E$_{\text{1g}}$ vibrations correspond to
a stretching in the $xz$ and $yz$ plane without changes along the
$z$ axis and therefore their frequencies vary with a slower
eccentricity dependence than the previous mode.  These rough
approximations are in agreement with the dependence observed in
figure \ref{FigSpheroid}.

Using perturbation theory,\citep{MariottoEL88} it is possible to
obtain more accurate expressions for the frequencies of these three
branches. The exact variations for $|R_z-R| \ll R$ are $\Omega
(1+4\beta/21)$, $\Omega (1-2\beta/21)$ and $\Omega (1-4\beta/21)$ for
the E$_{\text{1g}}$, E$_{\text{2g}}$ and A$_{\text{1g}}$ modes
respectively where $\beta = 2 (R_z-R)/(R_z+R)$ and $\Omega$ is the
frequency of modes S$_2^1$ for a spherical particle
having the same volume. Note that the length of the degenerate semi-axis
is constant in this work and therefore the volume varies
linearly with
$R_z$. Using $\Omega = \omega_{\text{S}_2^1} \sqrt[3]{R/R_z}$ results in
expressions which are in very good agreement with figure
\ref{FigSpheroid} close to $R_z = R$.

\begin{table}
\begin{tabular}{c|c|c|c}
  $i$  & $\nu$ (GHz) & i.r. & $\delta V$ \\
\hline
7-8 & 62.4 
 & E$_{\text{1u}}$
 & 0.0 \\
9 & 71.1
 & A$_{\text{2u}}$
 & 0.0 \\
10 & 90.8
 & A$_{\text{1g}}$
 & 0.4 \\
11-12 & 103.8 
 & E$_{\text{1g}}$
 & 0.0 \\
13 & 116.6
 & A$_{\text{2g}}$
 & 0.0 \\
14-15 & 127.0 
 & E$_{\text{1g}}$
 & 0.0 \\
16-17 & 136.3 
 & E$_{\text{2g}}$
 & 0.0 \\
\ldots & \ldots & \ldots & \ldots \\
34 & 185.2
 & A$_{\text{1g}}$
 & 0.2 \\
\ldots & \ldots & \ldots & \ldots \\
57 & 244.6
 & A$_{\text{1g}}$
 & 1.0 \\
\ldots & \ldots & \ldots & \ldots \\
64 & 253.3
 & A$_{\text{1g}}$
 & 0.0 \\
\ldots & \ldots & \ldots & \ldots \\
90 & 294.8
 & A$_{\text{1g}}$
 & 4.5 \\
\ldots & \ldots & \ldots & \ldots \\
110 & 318.8
 & A$_{\text{1g}}$
 & 0.5 \\
\ldots & \ldots & \ldots & \ldots \\
\end{tabular}
\caption{\label{TabSpheroid}Characteristics of modes of a spheroid
of dimensions $R=5$~nm, $R_z=10$~nm made from elastically isotropic
(\textit{i.e.} not mono-domain) Ag. $i$ is the mode index.}
\end{table}

Regarding the ``breathing'' mode, the picture gets very
complicated when $R_z/R$ differs significantly from 1. S$_0^1$ is
on the 4$^{\text{th}}$ A$_{\text{1g}}$ branch. Therefore the mode
of the spheroid with $R_z/R = 2$ on the same branch is mode
$i = 64$.  However, due to the mixing with neighboring
A$_{\text{1g}}$ branches, the mode with $i = 90$ is a better
candidate since its volume variation is much larger.  Thanks to
this additional property of the symmetric modes, it is
therefore possible to follow the breathing mode.  However,
because it is not possible to project onto the Lamb modes of an
isotropic sphere due to the different shapes, only the branches
and the anti-crossing patterns between branches having the same
irreducible representation can help tracking qualitatively the
other S and T modes.

These calculations are targetted at interpreting experimental
results on silver nanoparticles such as
nanocolumns.\cite{MargueritatNL06}
Compared to previous calculations using the FEMS method presented
before,\cite{MurrayPSSc06} the current approach enables a more complete
description of the different vibrations besides being faster and more
acurate.
In particular, the variation of volume is very efficient in
showing which modes should be observed by time-resolved pump-probe
experiments.\cite{BurginNL08}
It is interesting to note that some
works\cite{MargueritatNL06,BurginNL08} concern
aligned nanocolumns which results in interesting depolarization rules
for the Raman peaks.
For non-aligned nanoparticles, these rules are the same as those used
routinely for an ensemble of molecules, \textit{i.e.} all the Raman
active modes produce completely depolarized Raman peaks except for the
A$_{\text{1g}}$ vibrations which have a ``polarized'' scattering for
which the Raman peak is more intense when the polarizations of the
incident and scattered photons are parallel. For oriented
nanoparticles, this rule doesn't hold and the angles of the incident and
scattered photons with respect to the axis of symmetry of the
nanoparticles have to be taken into account.

Recent time-resolved pump-probe femtosecond experiments for
single gold nanoparticles\cite{TchebotarevaCPC09} also
demonstrate the need for such a model. In particular, the
observation of a peak close to the S$_2^1$ frequency is reported
for spheroids but not for spheres.  This is in agreement with the
features reported here for silver, namely that the
A$_{\text{1g}}$ mode coming from the S$_2^1$ mode has a non-zero
volume variation enabling it to be observed in such an experiment
unlike the S$_2^1$ mode of a sphere.  It is also worth
noting that the shape of the peak attributed to ``the breathing
mode'' in this work for dumbbells looks quite complicated.  Such
nano-objects have the same symmetry as spheroids. As a result, in
most cases there is no such thing as a ``breathing mode'' but
rather a set of A$_{\text{1g}}$ vibrations in a relatively narrow
frequency range having a significant volume variation.  This is
due to the fact that the A$_{\text{1g}}$ branch coming from the
S$_0^1$ vibration mixes with all the other A$_{\text{1g}}$
branches which come from all the S$_{\ell}$ vibrations with even
$\ell$.  Modeling such gold dumbbells is beyond the scope of this
paper, but doing so would enable a more detailed understanding of
the experimental results, especially for such single particle
measurements for which the external shape of the nanoparticles
can be obtained from SEM images.  The relatively large size of
these nanoparticles prevents them from being single-domain and
justifies the use of the isotropic elastic
approximation.

\subsubsection{anisotropic gold polyhedra}

Observation of faceted nanoparticles using electron microscopy is
quite common. Such facets can be thought of as a signature of the inner
crystal structure and therefore as an indication of the elastic
anisotropy.\cite{StephanidisPRB07} To quantify the importance of the
shape on the vibrations, we calculated the frequencies and irreducible
representations for some polyhedra and the results are presented in
table~\ref{TabPoly}. The crystal lattice is oriented with respect to the
shape so that the [100] planes correspond to the square faces for the
cuboctahedron and to the octagonal faces for the truncated
cuboctahedron. There is no lowering of symmetry associated
with these
shapes compared to the case of a mono-domain spherical gold
nanoparticle. The relevant point group is then D$_{\text{4h}}$. While
table~\ref{TabPoly} clearly shows that the lowest frequencies change
with the shape, the frequencies of the lowest E$_{\text{g}}$ and
T$_{\text{2g}}$ modes are hardly affected. Since the degeneracies of the
S$_0$ and S$_1$ modes are not lifted, all the modes which are observable
by Raman scattering, infrared absorption of time-resolved pump-probe
experiments are not sensitive to these changes of shape.

\begin{table}
\begin{tabular}{c|c|c|c}
i.r. & sphere & cuboctahedron & truncated cuboctahedron\\
    \hline
E$_{\text{g}}$  &  74.6 &  74.5 &  74.5\\
T$_{\text{2u}}$ &  76.9 &  73.4 &  75.5\\
A$_{\text{2g}}$ & 109.0 &  81.6 &  91.7\\
T$_{\text{1u}}$ & 111.2 & 108.7 & 110.7\\
E$_{\text{u}}$  & 114.1 & 102.3 & 107.9\\
T$_{\text{2g}}$ & 120.5 & 121.3 & 121.7\\
T$_{\text{2g}}$ & 140.2 & 128.0 & 133.8\\
\ldots &  \ldots & \ldots & \ldots\\
\end{tabular}
\caption{\label{TabPoly}Frequencies and irreducible representations
of the lowest frequency vibrations of elastically anisotropic gold
nanocrystals having different shapes. The volume of the different
nanocrystals is the same as that of a sphere of radius 5~nm.
The frequencies of the Raman active modes (E$_{\text{g}}$ and
T$_{\text{2g}}$) are almost unaffected by the shape.
See the text for the orientation of the crystal lattice with respect
to the polyedra.}
\end{table}

\section{Multiple-domain nanoparticles}

Several attempts have been made in the past either to fit low
frequency Raman spectra or to determine the size distribution
of the nanoparticles inside a sample using the shape of the
low-frequency Raman peak.  Both approaches always rely on the
validity of the isotropic model by Lamb and on the predominance
of the size distribution, the coupling with a surrounding matrix
and the electron-vibration coupling to fit the broadening of the
peaks.  However the distribution of internal structures of
multiple-domain nanoparticles also results in inhomogeneous
broadening.

It is in principle possible to model the vibrations of a
multiple-domain nanoparticle using the
numerical method of Visscher \textit{et al.}.
However this requires a complete description of the position of
the domain boundaries and the orientations of the crystal lattice.
For ensemble measurements with the nanoparticles having a variety
of different internal structure, a lot of calculations would be
required. Otherwise, using such an approach for a single
internal structure is justified only if all the studied
nanoparticles are identical for ensemble measurements or if the
inner structure of a nanoparticle studied in a single particle
measurement is perfectly known. Due to these latter two
conditions having never been met until now and also to the
additional complexity of modelling a multiple-domain
nanoparticle, we suggest using the isotropic approximation to describe
an ensemble of multiple-domain nanoparticles. What this means is that no
nanoparticle behaves exactly as an isotropic nanoparticle, but
the isotropic approximation gives an average value due to the
nanoparticles having essentially random domain structures.
Ensemble measurements should therefore show features associated
with these average frequencies with some inhomogeneous width due
to the inhomogeneous distribution of domains inside
the population of nanoparticles.

We calculated the vibrations of an icosahedron made of gold and
silver assuming the isotropic approximation to be valid in that
case.  The corresponding point group for such a nanoparticle is
I$_{\text{h}}$.  It is interesting to note that in this case
there is no degeneracy lifting for the modes S$_0$, S$_1$ and
S$_2$ whose irreducible representation in the new system is
A$_{\text{g}}$, T$_{\text{1u}}$ and H$_{\text{g}}$ respectively.
As a result, there is hardly any frequency difference compared
to the case of a sphere having the same volume.  A real
icosahedron made of an anisotropic material but having the same
symmetry due to the presence of twins would have no
degeneracy lifting either for the same modes and we expect almost
the same frequencies too.

\section{Conclusion}

We have presented mode frequencies and irreducible
representations for homogeneous continuum nanoparticles using
a standard numerical method which can handle arbitrary shape
and anisotropic elasticity. The classification by irreducible
representation makes it possible to label many modes as either
Raman or infrared inactive. We have been able to go beyond this
to provide some tools to make qualitive estimates of the Raman
intensity of potentially Raman active modes. These tools are
well-suited for the interpretation of experimental results
obtained with vibrational spectroscopies.\cite{PortalesPNAS08,SaviotPRB08} 

This approach fills a large gap in current works since up to now
it was necessary to choose between a simplified spherical
isotropic model where the underlying physics was simple enough
but where the accuracy was challenged by recent experimental
results or a numerical approach which provides accurate
frequencies but which is inefficient in practice due to the lack
of tools to distinguish the relevant vibrations without complex
simulations of spectra.

It is clear that additional theoretical work is required in order
to quantitatively predict Raman intensities for elastically
anisotropic modes of spheres. However, the labeling of modes via
their Lamb mode parentage by continuous variation of the elastic
isotropy and projections for spherical nanoparticles, is a
powerful descriptive and semi-quantitative tool for understanding
what is going on with elastic anisotropy.
The significant frequency splittings obtained in this work call into
question the validity of the isotropic approximation for the case of
multiple-domain nanoparticles. Such systems are very relevant
experimentally but their vibrations remain largely unaddressed.

\begin{acknowledgments}

LS acknowledges Professor Eug\`ene Duval for stimulating
discussions and comments. DBM acknowledges support from NSERC
of Canada.

\end{acknowledgments}

\bibliography{sea}

\end{document}